\documentclass[runningheads]{llncs}
\usepackage{graphicx}
\usepackage{amsmath}
\usepackage{amsfonts,amssymb}
\usepackage{booktabs}
\usepackage{multirow}
\usepackage{bbding}
\usepackage{makecell}
\usepackage{float}
\usepackage{bbm}
\usepackage{hyperref}
\PassOptionsToPackage{table,xcdraw}{xcolor}
\usepackage[misc]{ifsym}
\begin{document}
\pagestyle{empty}
\title{To Recommend or Not: Recommendability Identification in Conversations with Pre-trained Language Models}
%
\titlerunning{To Recommend or Not}
%
\author{Zhefan Wang\inst{1,2} \and
Weizhi Ma\inst{3}\Letter \and
Min Zhang\inst{2,1}\Letter}
%
\authorrunning{Z. Wang et al.}
%
\institute{Department of Computer Science and Technology, Tsinghua University, Beijing 100084, China \and
Quan Cheng Laboratory \and
Institute for AI Industry Research (AIR), Tsinghua University \email{wzf23@mails.tsinghua.edu.cn, \{mawz,z-m\}@tsinghua.edu.cn}}
\maketitle
\begin{abstract}



Most current recommender systems primarily focus on what to recommend, assuming users always require personalized recommendations. However, with the widely spread of ChatGPT and other chatbots, a more crucial problem in the context of conversational systems is how to minimize user disruption when we provide recommendation services for users. 
While previous research has extensively explored different user intents in dialogue systems, fewer efforts are made to investigate whether recommendations should be provided. 
In this paper, we formally define the recommendability identification problem, which aims to determine whether recommendations are necessary in a specific scenario.
First, we propose and define the recommendability identification task, which investigates the need for recommendations in the current conversational context.
A new dataset is constructed.
Subsequently, we discuss and evaluate the feasibility of leveraging pre-trained language models (PLMs) for recommendability identification.
Finally, through comparative experiments
, we demonstrate that directly employing PLMs with zero-shot results falls short of meeting the task requirements.
Besides, fine-tuning or utilizing soft prompt techniques yields comparable results to traditional classification methods.
Our work is the first to study recommendability before recommendation and provides preliminary ways to make it a fundamental component of the future recommendation system.\footnote{Code is available at \href{https://github.com/wzf2000/Recommendability\_DASFAA2024}{https://github.com/wzf2000/Recommendability\_DASFAA2024}.}

\end{abstract}

\section{Introduction}

In recent years, the rapid development of large-scale language models (LLMs) supported by massive model parameters and data has led to significant advancements in multilingual dialogue systems.
Conversational interfaces are expected to become the primary mode of human-computer interaction.
However, existing dialogue systems predominantly focus on answering user questions and fulfilling information requests, representing a passive interaction.
In real-life scenarios, the problem of information overload necessitates proactive information filtering, often facilitated by recommender systems.
As for traditional dialog systems, proactively identifying and satisfying user needs is a desired capability.
The lack of this capability will also limit overall user satisfaction.

Prior research has explored the integration of recommender and dialogue systems, such as conversational recommender systems and dialogue-based search and recommendation.
However, these efforts are often similar to traditional recommender systems, assuming users actually need the recommendation results.
Moreover, the research primarily concentrates on improving the recommendation performance by investigating what items to recommend.
Nevertheless, in more general conversational contexts, users may have diverse needs beyond simple recommendation-seeking.
For instance, in a mobile dialogue retrieval scenario, the system can provide recommendations and searches specific to mobile shopping.
However, in the broader context of this study, making recommendations without careful consideration may disrupt the user's experience.
A dialogue system should determine when to offer recommendations during a conversation and when to address the user's queries solely.
While prior research has extensively studied user intents in dialogue systems, there has been limited investigation into ``whether to recommend or not."

This work aims to study the recognition of recommendability, which explores whether recommendations are necessary for general dialogue processes and identifies suitable moments to give recommendations.
It is worth noting that recommendability is a dynamic characteristic that evolves throughout a conversation with the same user.
For instance, once a user obtains satisfactory recommendation results, the system should adjust its identification of recommendability in real time.
Hence, our research focuses on utterance-level recommendability identification rather than conversation-level analysis, which facilitates the evaluation of the system's real-time dynamic identification capabilities.

In this work, we propose and define the task of recommendability identification based on conversational scenarios.
To our knowledge, this task has not been previously proposed by other researchers.
Due to the novelty of this task, there needs to be more publicly available dialogue datasets suitable for recommendability annotations.
Therefore, we manually annotated the recommendability of sampled user dialogues from a real-world e-commerce customer service dialogue dataset, JDDC2.1, and created a processed JDDCRec dataset.
Considering those pre-trained language models (PLMs) are widely adopted in dialogue scenarios, we primarily employ PLMs and prompt learning techniques for experimental evaluation.
The main contributions of this work can be summarized as follows:
\begin{itemize}
    \item[$\bullet$] We propose and formally define the recommendability identification task, which is crucial for applying recommendations on real conversational systems. We also construct the JDDCRec dataset for the evaluation of this task.
    \item[$\bullet$] We discuss and evaluate the potential of using PLMs and prompt-based methods, such as zero-shot prompt evaluation, prompt learning, and soft prompt tuning, for recommendability identification.
    \item[$\bullet$] Experimental results on existing dialogue-based recommendation datasets and annotated JDDCRec demonstrate the potential and feasibility of utilizing PLMs and prompt-based methods for recommendability identification.
\end{itemize}
\section{Related Work}

\subsection{Conversational Search and Recommendation}


Recently, the rise of pre-trained language models (PLMs) in NLP, especially LLMs, has made human-machine dialogues more popular in general scenarios. Integrating dialogue systems is a basic form of future development in search and recommendation.

Some studies\cite{culpepper2018research,vakulenko2019knowledge} view conversational search and recommendation as task-oriented dialogue, aiming to meet user's information needs.
Compared to traditional dialogue systems, conversational search and recommendation can help users find products and items that align with their preferences and allow users to provide immediate feedback directly.
Tran et al.\cite{tran2020deep} specifically summarize the critical components of the conversational recommender system (CRS), including understanding the user's intention, providing personalized recommendations, and developing a suitable switch mechanism.

With the widespread application of LLMs, many researchers have attempted to integrate LLMs with Conversational Search Systems (CSS) and CRS.
Friedman et al.\cite{friedman2023leveraging} propose leveraging LLMs to build an end-to-end large-scale CRS roadmap.
Mao et al.\cite{mao2023large} introduce a framework called LLMCS (LLM-based Conversational Search), which leverages LLMs to perform few-shot conversational query rewriting for conversational search.

However, most existing research on CSS and CRS focuses on relatively single-domain dialogue scenarios, often assuming user's purchase intent in e-commerce. In this work, we aim to study more mixed dialogue scenarios, where
the topics and domains of the dialogue are more open-ended. Identifying user intent becomes crucial in such scenarios, as the system's responses directly impact user experience and satisfaction.

\subsection{Intent Detection for Dialogue System}


Intent detection is a crucial task in dialogue systems, aiming to accurately understand the intentions expressed by users in conversations and provide appropriate responses accordingly.

Some commonly used methods for intent detection include statistical approaches such as Support Vector Machine (SVM)\cite{haffner2003optimizing} and Naive Bayes\cite{mccallum1998comparison},
and some deep neural network methods, such as CNN\cite{hashemi2016query}, and RNN\cite{kim2016intent}, etc.
Compared to traditional statistical methods, deep learning approaches can better capture the semantics in dialogues, resulting in significant improvements in performance on intent detection.

In recent years, PLMs, such as BERT\cite{devlin2018bert}
and GPT\cite{radford2018improving}, which have learned rich language representations through unsupervised pre-training on large-scale textual corpora, have gained widespread attention.
Researchers have attempted to apply PLMs to intent detection and achieved excellent results\cite{castellucci2019multi}.
Compared to traditional deep learning methods, they excel in learning deep language representations from large-scale unsupervised data, exhibiting more robust performance across many downstream tasks.

In this work, recommendability cannot be treated purely as a user's intent but should be regarded as opportunities to recommend within the conversation. 
A user without information-seeking intent might be interested in the recommendation results provided by the system.
Conversely, users looking to retrieve relevant information may feel disturbed by the system's abrupt intervention in recommendations.

\subsection{Recommendation with Pre-trained Language Model}



We have witnessed the growing power of PLMs in language modeling in recent years. Many studies have attempted to leverage the robust capabilities of PLMs for recommendation tasks. Some works encode the textual features of users and items, such as news articles and user reviews, using PLMs as encoders\cite{zhang2021unbert,jia2021rmbert,wu2021empowering}, benefiting from the powerful text comprehension abilities of PLMs to enhance recommendations. Other works directly utilize PLMs as backbones for recommendation\cite{zhang2021language,li2023personalized,geng2022recommendation,cui2022m6}. For instance, Zhang et al. (2021) transformed the recommendation task into a language modeling task to address the issue of zero-shot recommendations, while Li et al. (2023) attempted to convert user and item ID information into continuous or discrete token representations, enabling PLMs to learn the recommendation task based on these representations. Geng et al. (2022) and Cui et al. (2022) focused on different tasks in various domains of recommendation systems, designing a unified framework to adapt to a wide range of possible recommendation scenarios.

However, the works mentioned above assume that the system performs recommendations, while recommendability identification is a preliminary step before providing specific recommendation results. In other words, meaningful recommendation results can only be derived after ensuring that the system's recommendations will not disturb the user, highlighting the significance of recommendability identification. Therefore, this study addresses the initial step in most recommendation system research. A certain level of understanding of the dialogue text is required in general dialogue scenarios. Hence, this paper explores the use of PLMs for recommendability identification.
\section{Definition of Recommendability Identification Task}

Let $u_i$ be the $i$-th utterance during the conversation, and $r_i \in \{0, 1\}$ represents whether giving a recommendation is appropriate when given the first $i - 1$ utterance (1 indicating appropriate while 0 indicating inappropriate; $r_i$ is set to 0 for every user utterance $u_i$). Let $\mathcal{P}$ represent the user's profile, including attributes such as age, gender, and interests, while $\mathcal{C}$ represents additional contextual information during the conversation, such as time or locations. The goal of recommendability identification is to predict whether a recommendation is appropriate for each system response timing, i.e.,
\begin{equation}
    \hat{r}_i = f_\Theta\left(u_1, u_2, \cdots, u_{i - 1}\left[, r_1, r_2, \cdots, r_{i - 1}, \mathcal{P}, \mathcal{C}\right] \right) \in \{0, 1\},
\end{equation}
where $\Theta$ denotes the parameters of the model. Inputs enclosed in $[\dots]$ are optional, as their importance may vary depending on the specific application scenario.

\section{Recommendability Identification with Pre-trained Language Model}

\begin{figure}[H]
    \centering
    \includegraphics[width=1\linewidth]{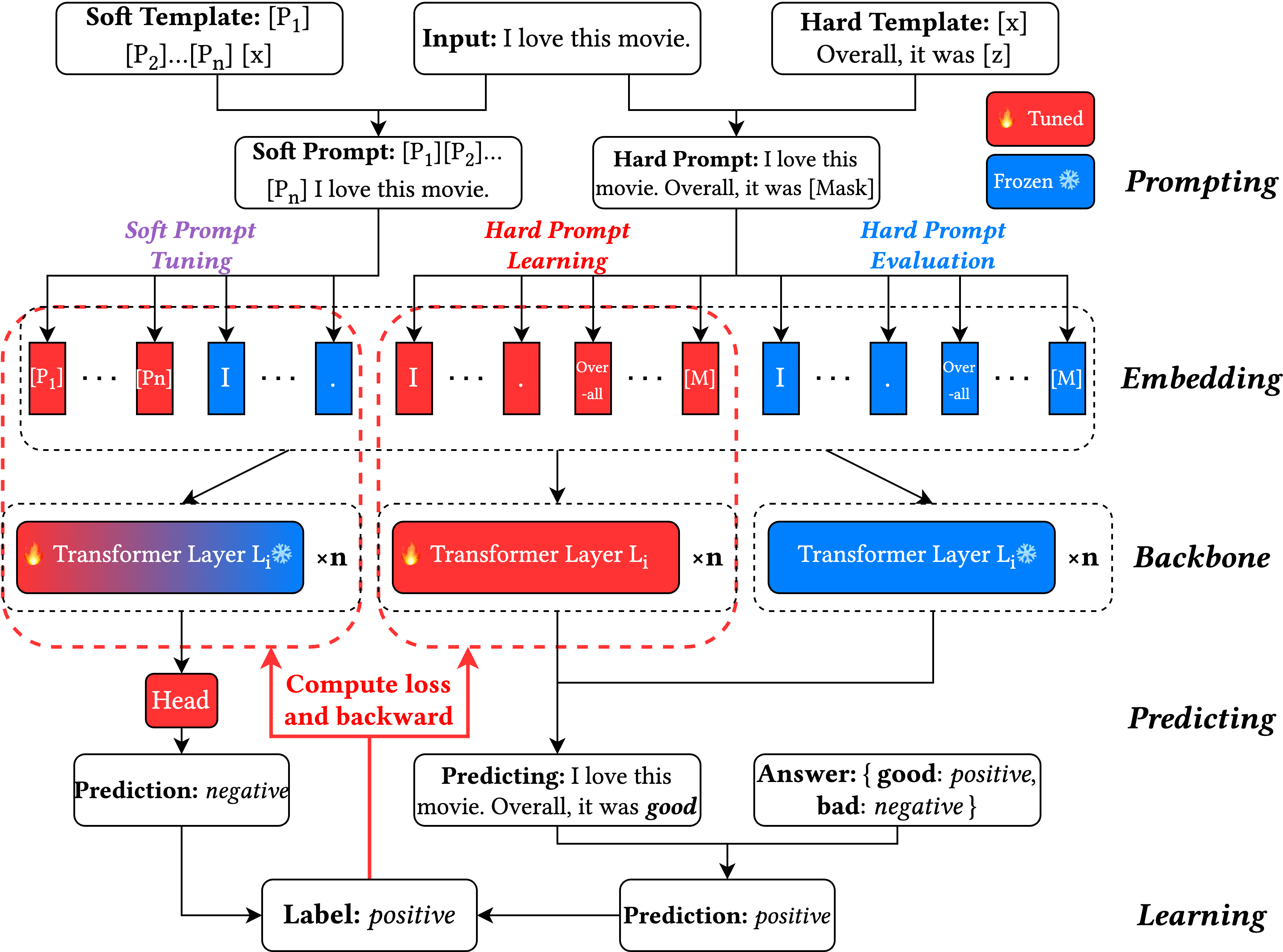}
    \caption{Flow comparison of the three methods. The red indicates the parameters involved in the training, and the blue indicates frozen parameters.}
    \label{fig:method}
\end{figure}
We compare the three methods mentioned in this section in Fig.~\ref{fig:method}.
\subsection{Zero-Shot Validation with Manual Template}



With powerful PLMs like BERT \cite{devlin2018bert}, RoBERTa \cite{liu2019roberta}, and DeBERTa \cite{radford2019language}, prompting the input text sequence with manual template\cite{liu2023pre} can convert various downstream tasks into the similar form.
As illustrated in Fig.~\ref{fig:method}, taking the recommendability Identification task for e-commerce customer service conversation as an example, we manually construct a template as follows:

\textit{Assuming that you are an intelligent e-commerce customer service, you can intelligently determine the needs of customers in the process of communicating with customers, and give recommendations when needed.The following is the dialogue history between you and a customer:} \verb|[x]| \textit{You will choose? Options: 0: no recommendation; 1: recommendation. Answer:} \verb|[Mask]|

\verb|[x]| will be replaced by the historical utterances, and the PLM will be required to predict the token at \verb|[Mask]|.
The optional space of the token consists of $\{0, 1\}$, indicating non-recommendable and recommendable, respectively.

According to Brown et al. \cite{brown2020language} and Wei et al. \cite{wei2021finetuned}, large-scale PLMs can obtain powerful few-shot and zero-shot capabilities through instruction tuning.
Therefore, we evaluated the ability of PLMs with different scales on the recommendability identification task under zero-shot conditions.

\subsection{Prompt Learning with Manual Template}



To test the limits of the PLMs' capabilities on the recommendability identification task, we further required the PLMs to predict the probability distribution of the token at \verb|[Mask]|. As shown in the part of hard Prompt learning in Fig.~\ref{fig:method}, all PLMs' parameters were fine-tuned using binary cross-entropy loss:
\begin{equation}
    \mathcal{L}_{\mathrm{CE}}(\tilde{\pmb{P}}, \pmb{y}) = -\frac{1}{N} \left[\sum_{i = 1}^N \left(\mathbbm{1}_{y_i = 1} \ln \tilde{P}_{i, 1} + \mathbbm{1}_{y_i = 0} \ln \tilde{P}_{i, 0}\right) \right],
    \label{eq:ce}
\end{equation}
where $\tilde{P}_{i, c}$ denotes the probability of $c$-th category ($c \in \{0, 1\}$) for the $i$-th data sample, $y_{i}$ denotes the one-hot encoded representation of the ground truth data category for the $i$-th sample.

\subsection{Soft Prompt Tuning}
+

In addition to hard templates composed of natural language, some researchers\cite{liu2021gpt} has also proposed to directly use continuous word vector representations instead of manual templates, which is known as \emph{soft prompt}.
For the original template in the form of "\verb|[x]|\verb|[Prompt]|\verb|[Mask]|", after mapping through the embedding layer, the model's input will be:
$$
[\pmb{e}(\pmb{x}), \pmb{h}_0, \pmb{h}_1, \cdots, \pmb{h}_m, \pmb{e}(``[\mathrm{Mask}]")].
$$

Furthermore, Liu et al. \cite{liu2021p} proposed P-Tuning V2 to alleviate the poor performance of P-Tuning on small-scale PLMs.
P-Tuning V2 introduces similar trainable continuous prompt parameters before the first layer of the encoder and multiple layers of the encoder.
As illustrated in Fig.~\ref{fig:method}, only the newly introduced soft prompt parameters are involved in training.
The specific computational process is outlined below:
\begin{equation}
    \begin{aligned}
        \tilde{\pmb{e}}^0 & = \left[\pmb{h}^0_0, \pmb{h}^0_1, \cdots, \pmb{h}^0_m, \pmb{e}(\pmb{x})\right], \\
        \pmb{e}^i & = 
        \begin{cases}
            f_{\mathrm{encoder}, i}\left(\tilde{\pmb{e}}^{i - 1}\right), & \text{Add soft prompt for $i$-th layer} \\
            f_{\mathrm{encoder}, i}\left(\pmb{e}^{i - 1}\right), & \text{Otherwise}
        \end{cases}, \\
        \tilde{\pmb{e}}^i & = \left[\pmb{h}^i_0, \pmb{h}^i_1, \cdots, \pmb{h}^i_m, \pmb{e}^i_{m + 1:}\right].
    \end{aligned}
    \label{eq:pt2}
\end{equation}

\section{Experiments}

\subsection{Construction of JDDCRec Dataset}

\subsubsection{Dataset Construction Strategy}

According to the definition of the recommendability identification task, we selected the JDDC2.1\cite{zhao2022jddc} dataset as the foundation for manual annotation.
JDDC2.1 is a large-scale multimodal multi-turn dialogue dataset collected from a mainstream Chinese E-commerce platform, containing 246K sessions, 3M utterances, and 507K images.

The cost of thoroughly labeling the JDDC2.1 dataset is significant since the identification of recommendability relies on a high level of understanding of conversation history.
Besides, the ratio of recommendability in the general conversation scenarios is meager.
Therefore, we first performed several rounds of rule-based filtering on JDDC2.1, e.g., dialogs containing information about returns, logistics, etc.
Contexts containing such information have little possibility of recommendability.
Based on this, we filtered JDDC2.1 to 22.3\% of its original size.

We sampled 1000 data from the filtered dataset for final manual labeling (8:1:1 ratio of training, validation, and testing).
For labeling, we employed three professional annotators.
The annotators were required to browse each conversation and determine whether there was an opportunity to recommend at the utterance level according to our criteria.
To maintain annotation consistency, we provided many examples, including detailed analytical explanations for several ambiguous cases.

\begin{table}
\vspace{-1.0em}
\centering
\caption{Dataset characteristics for DuRecDial 2.0 and JDDCRec. Train/Dev/Test indicates the number of train/development/test set samples.}
\label{tab:datasetsize}
\begin{tabular}{ccccc}
  \toprule
  \textbf{Dataset} & \textbf{Train} & \textbf{Dev} & \textbf{Test} & \textbf{Positive ratio}  \\
  \midrule
  DuRecDial 2.0 & 33079 & 4801 & 10456 & $15.2\%$ \\
  JDDCRec & 1300 & 136 & 159 & $29.5\%$ \\
  \bottomrule
\end{tabular}
\vspace{-1.0em}
\end{table}
\subsubsection{Dataset Characteristics}



\begin{figure}[H]
    \centering
    \includegraphics[width=1\linewidth]{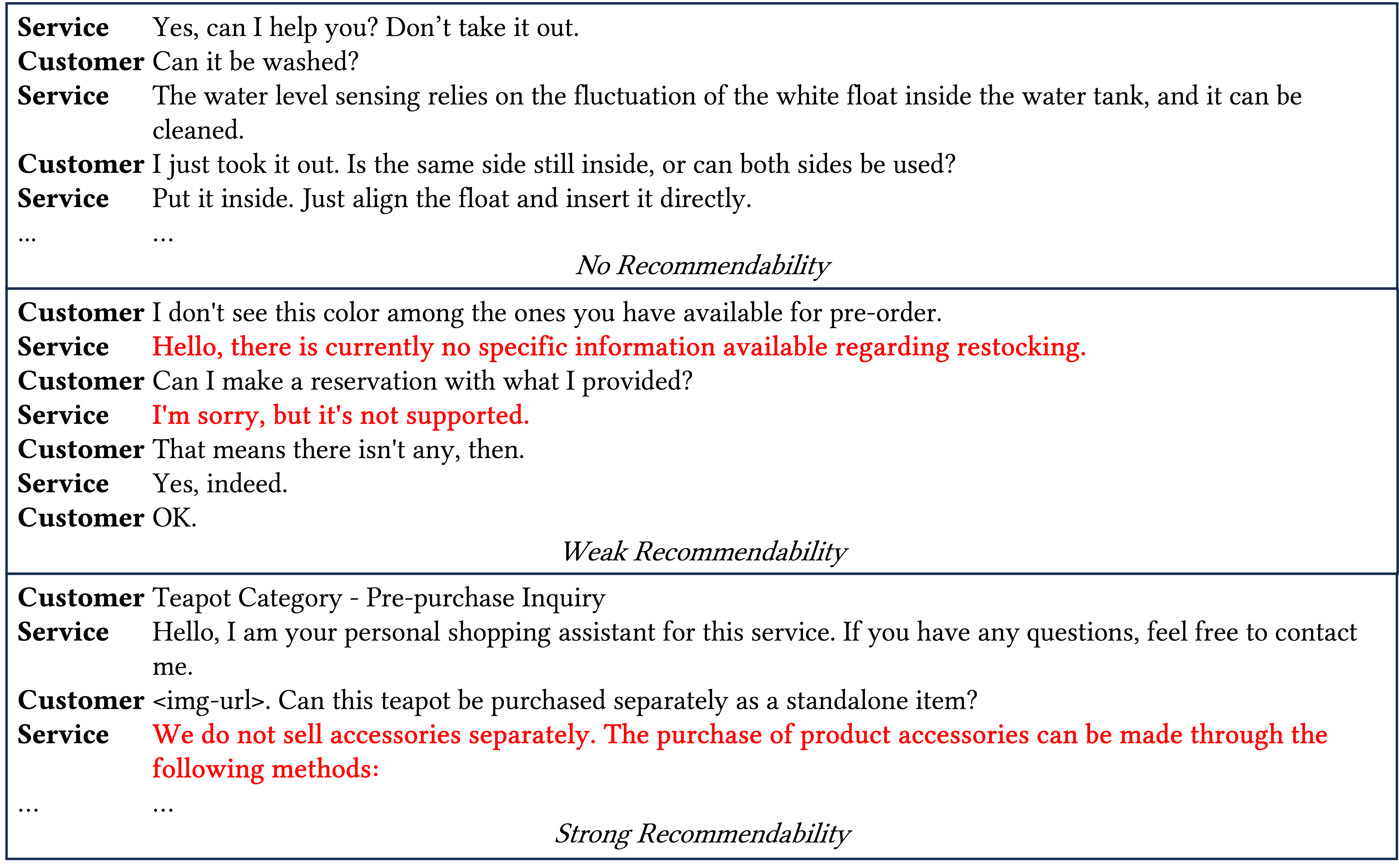}
    \caption{Some conversation examples from JDDCRec. Bold red indicates that there is a possibility of recommendation. Here, we distinguish between the existence and the strength of the recommendability.}
    \label{fig:jddc_sample}
\end{figure}

After the above construction process, we obtained 1000 conversations containing utterance-level annotations.
As a result of our analysis, the filtered JDDC2.1 dataset (i.e., the JDDCRec dataset) contains 220 conversations and 471 utterances with recommendability annotations.
We give some samples of the data in JDDCRec in Fig.~\ref{fig:jddc_sample}.

Considering the imbalance of positive and negative cases in JDDCRec, we retained only the last of the consecutive negative examples for each conversation, encompassing a more comprehensive representation of the entire conversation.
Following these preprocessing steps, the resulting JDDCRec dataset used for experimentation is summarized in Table~\ref{tab:datasetsize}.

\subsection{Experimental Settings}

\subsubsection{Datasets.}
In addition to the JDDCRec dataset mentioned above, we also use another dataset, DuRecDial 2.0\cite{liu2021durecdial}.
DuRecDial 2.0 is the first publicly available bilingual parallel dataset for conversational recommendation.
It encompasses various domains, including movies, celebrities, cuisine, music, weather, and points of interest (POI), comprising 8241 conversations.
The DuRecDial 2.0 dataset contains utterance-level topic annotations used as recommendability annotations in our experiments (based on whether the topic contains ``recommendation" or not).

The original DuRecDial 2.0 dataset always had multiple rounds of the same topic.
However, we tended to focus more on where the first recommendability appeared
Therefore, we retained only the first positive example for the same topics in our data processing.

After the above processing, the size of the DuRecDial2.0 dataset can be found in Table~\ref{tab:datasetsize}.

\subsubsection{Baselines.}We compared the three types of methods mentioned in this work with several baselines as follows,
\begin{itemize}
    \item[$\bullet$] ConvBERT\cite{zhou2020towards}: Conversation BERT. It is a topic prediction model for a topic-guided conversational recommender system. It simply concatenates all the historical utterances (with \verb|[SEP]| tokens for separating) and encodes them with the BERT-base\cite{devlin2018bert} model. To get a probability for recommendability identification, we apply a linear transformation to the obtained representation.
    \item[$\bullet$] ConvDeBERTa/ConvRoBERTa: Similar to ConvBERT, but uses DeBERTa-base\cite{radford2019language} and RoBERTa-base\cite{liu2019roberta} as the encoding model.
\end{itemize}


\subsubsection{Metrics.}We report accuracy, precision, recall, and F1 in the main experiment.

\subsubsection{Implement details.}Our experiments are conducted through OpenPrompt (Ding et al. \cite{ding2021openprompt}) and CRSLab (hou et al. \cite{zhou2021crslab}).
We used the Fengshenbang Chinese Language Pre-training Model Open Source Project \cite{fengshenbang} as a valuable resource to find suitable Chinese pre-trained language models in our experiments.

\begin{table*}[htb]
\caption{Experimental results of three different methods and baseline methods on the JDDCRec dataset. The boldface indicates the best of the specific method type, and underlining indicates the best of all methods.}
\label{tab:jddc}
\resizebox{\linewidth}{!}{
\begin{tabular}{c|c|cccc}
\toprule
Method Type & Model & Accuracy  & Precision & Recall & F1 \\
\midrule
\multirow{3}{*}{baseline}
& ConvBERT & 0.8252 & 0.7058 & \textbf{0.6913} & 0.6922 \\
& ConvDeBERTa & 0.8226 & 0.7323 & 0.6304 & 0.6642 \\
& ConvRoBERTa & \textbf{0.8516} & \underline{\textbf{0.7891}} & 0.6696 & \textbf{0.7205} \\
\midrule
\multirowcell{5}{Zero-shot \\ Prompt Evaluation}
& BERT-base & 0.4277 & 0.2581 & 0.5217 & 0.3453 \\
& RoBERTa-base & 0.2893 & 0.2893 & \underline{\textbf{1.0000}} & 0.4488 \\
& RoBERTa-large & 0.2893 & 0.2893 & \underline{\textbf{1.0000}} & 0.4488 \\
& ChatGLM-6B & 0.5597 & 0.3723 & 0.7609 & \textbf{0.5000} \\
& ChatGLM2-6B & \textbf{0.7170} & \textbf{0.5263} & 0.2174 & 0.3077 \\
\midrule
\multirow{3}{*}{Hard Prompt Learning}
& BERT-base & 0.8302 & 0.6922 & \textbf{0.7435} & \textbf{0.7168} \\
& RoBERTa-base & 0.8365 & 0.7796 & 0.6131 & 0.6837 \\
& RoBERTa-large & \textbf{0.8377} & \textbf{0.7857} & 0.6130 & 0.6861 \\
\midrule
\multirow{3}{*}{Soft Prompt Tuning}
& BERT-base & 0.8164 & 0.6679 & 0.7087 & 0.6835 \\
& RoBERTa-base & 0.8176 & 0.7049 & 0.6522 & 0.6709 \\
& RoBERTa-large & \underline{\textbf{0.8654}} & \textbf{0.7336} & \textbf{0.8522} & \underline{\textbf{0.7856}} \\
\bottomrule
\end{tabular}
}
\end{table*}

\subsection{Overall Evaluation Results}

The overall performance for two datasets is shown in Table~\ref{tab:jddc}, Table~\ref{tab:durecdial-en}, and Table~\ref{tab:durecdial-zh}, which contains four metrics for the recommendability identification task.
We divide each table into four parts:
The first part is the baseline methods.
The second and third parts are the method of zero-shot evaluation and classic prompt learning using the manual template.
The last part shows the results of soft prompt tuning, which uses the P-Tuning V2 method.

\begin{table*}[htb]
\centering
\caption{Experimental results of three different methods and baseline methods on DuRecDial 2.0 English dataset.}
\label{tab:durecdial-en}
\begin{tabular}{c|c|cccc}
\toprule
Method Type & Model & Accuracy  & Precision & Recall & F1 \\
\midrule
\multirow{3}{*}{baseline}
& ConvBERT & 0.9779 & 0.9550 & 0.9053 & 0.9288 \\
& ConvDeBERTa & \underline{\textbf{0.9797}} & \textbf{0.9618} & 0.9129 & 0.9344 \\
& ConvRoBERTa & 0.9795 & 0.9549 & \textbf{0.9184} & \underline{\textbf{0.9345}} \\
\midrule
\multirowcell{10}{Zero-shot \\ Prompt Evaluation}
& BERT-base & 0.7244 & 0.0444 & 0.0353 & 0.0393 \\
& BERT-large & 0.5484 & 0.1706 & 0.4731 & 0.2508 \\
& RoBERTa-base & 0.4717 & 0.1925 & 0.7222 & \textbf{0.3040} \\
& RoBERTa-large & 0.1600 & 0.1598 & \underline{\textbf{1.0000}} & 0.2756 \\
& GPT2-base & 0.1622 & 0.1600 & 0.9994 & 0.2758 \\
& GPT2-medium & 0.2091 & 0.1679 & 0.9988 & 0.2875 \\
& GPT2-large & 0.6812 & 0.1761 & 0.2707 & 0.2134 \\
& GPT2-XL & \textbf{0.8209} & 0.1988 & 0.0401 & 0.0667 \\
& ChatGLM-6B & 0.8168 & \textbf{0.3016} & 0.1120 & 0.1633 \\
& ChatGLM2-6B & 0.8183 & 0.0472 & 0.0072 & 0.0125 \\
\midrule
\multirow{6}{*}{Hard Prompt Learning}
& BERT-base & 0.9777 & 0.9633 & \textbf{0.8969} & 0.9277 \\
& BERT-large & 0.9765 & 0.9691 & 0.8822 & 0.9226 \\
& RoBERTa-base & \textbf{0.9791} & 0.9732 & 0.8950 & \textbf{0.9315} \\
& RoBERTa-large & 0.9783 & 0.9677 & 0.8959 & 0.9290 \\
& GPT2-base & 0.9780 & \textbf{0.9865} & 0.8740 & 0.9268 \\
& GPT2-medium & 0.9778 & 0.9810 & 0.8783 & 0.9266 \\
\midrule
\multirow{5}{*}{Soft Prompt Tuning}
& BERT-base & 0.9767 & 0.9730 & \textbf{0.8787} & 0.9232 \\
& BERT-large & 0.9772 & 0.9765 & 0.8786 & 0.9248 \\
& RoBERTa-base & 0.9780 & 0.9855 & 0.8753 & 0.9271 \\
& RoBERTa-large & 0.9780 & 0.9853 & 0.8752 & 0.9270 \\
& DeBERTa-base & \textbf{0.9783} & \underline{\textbf{0.9866}} & 0.8761 & \textbf{0.9280} \\
\bottomrule
\end{tabular}
\end{table*}

\begin{table*}[htb]
\centering
\caption{Experimental results of three different methods and baseline methods on DuRecDial 2.0 Chinese dataset.}
\label{tab:durecdial-zh}
\begin{tabular}{c|c|cccc}
\toprule
Method Type & Model & Accuracy  & Precision & Recall & F1 \\
\midrule
\multirow{3}{*}{baseline}
& ConvBERT & 0.9754 & 0.9582 & 0.8885 & 0.9195 \\
& ConvDeBERTa & \underline{\textbf{0.9806}} & 0.9554 & \textbf{0.9252} & \underline{\textbf{0.9379}} \\
& ConvRoBERTa & 0.9800 & \textbf{0.9585} & 0.9176 & 0.9356 \\
\midrule
\multirowcell{5}{Zero-shot \\ Prompt Evaluation}
& BERT-base & 0.2444 & 0.1583 & 0.8641 & 0.2676 \\
& RoBERTa-base & 0.6537 & 0.1586 & 0.2713 & 0.2002 \\
& RoBERTa-large & 0.1597 & 0.1597 & \underline{\textbf{1.0000}} & 0.2754 \\
& ChatGLM-6B & \textbf{0.6982} & 0.1538 & 0.1976 & 0.1730 \\
& ChatGLM2-6B & 0.1898 & \textbf{0.1618} & 0.9743 & \textbf{0.2775} \\
\midrule
\multirow{3}{*}{Hard Prompt Learning}
& BERT-base & 0.9775 & 0.9684 & 0.8893 & 0.9267 \\
& RoBERTa-base & \textbf{0.9787} & 0.9732 & \textbf{0.8920} & \textbf{0.9304} \\
& RoBERTa-large & 0.9769 & \textbf{0.9795} & 0.8739 & 0.9237 \\
\midrule
\multirow{3}{*}{Soft Prompt Tuning}
& BERT-base & 0.9767 & 0.9663 & \textbf{0.8862} & 0.9240 \\
& RoBERTa-base & \textbf{0.9783} & 0.9782 & 0.8838 & \textbf{0.9285} \\
& RoBERTa-large & 0.9776 & \underline{\textbf{0.9899}} & 0.8686 & 0.9253 \\
\bottomrule
\end{tabular}
\end{table*}


From the results, we have several observations:

First, the three baseline methods achieved excellent performance on both datasets.
The above results indicate the effectiveness of the traditional encoder plus classifier approach in the recommendability identification task.

Second, the zero-shot prompt evaluation method performed poorly on both datasets.
Only a few models based on larger-scale backbones, such as GPT2-XL and ChatGLM-6B, showed good accuracy but performed poorly in the F1 score.
These observations suggest that using simple manual templates to prompt PLMs does not directly lead to better performance in recommendability identification.

Third, the hard prompt learning and soft prompt tuning methods showed significant improvements compared to the second part and performed comparably to the baseline methods across all four metrics.
The fourth part further indicates the possibility of directly eliciting PLMs' identification ability by constructing templates, as soft prompt tuning only introduces additional prefix prompt parameters and freezes the backbone's original parameters during the learning process.


\subsection{Further Analyses}

\subsubsection{Few-shot Performace}

\begin{figure*}
    \centering
    \includegraphics[width=1\linewidth]{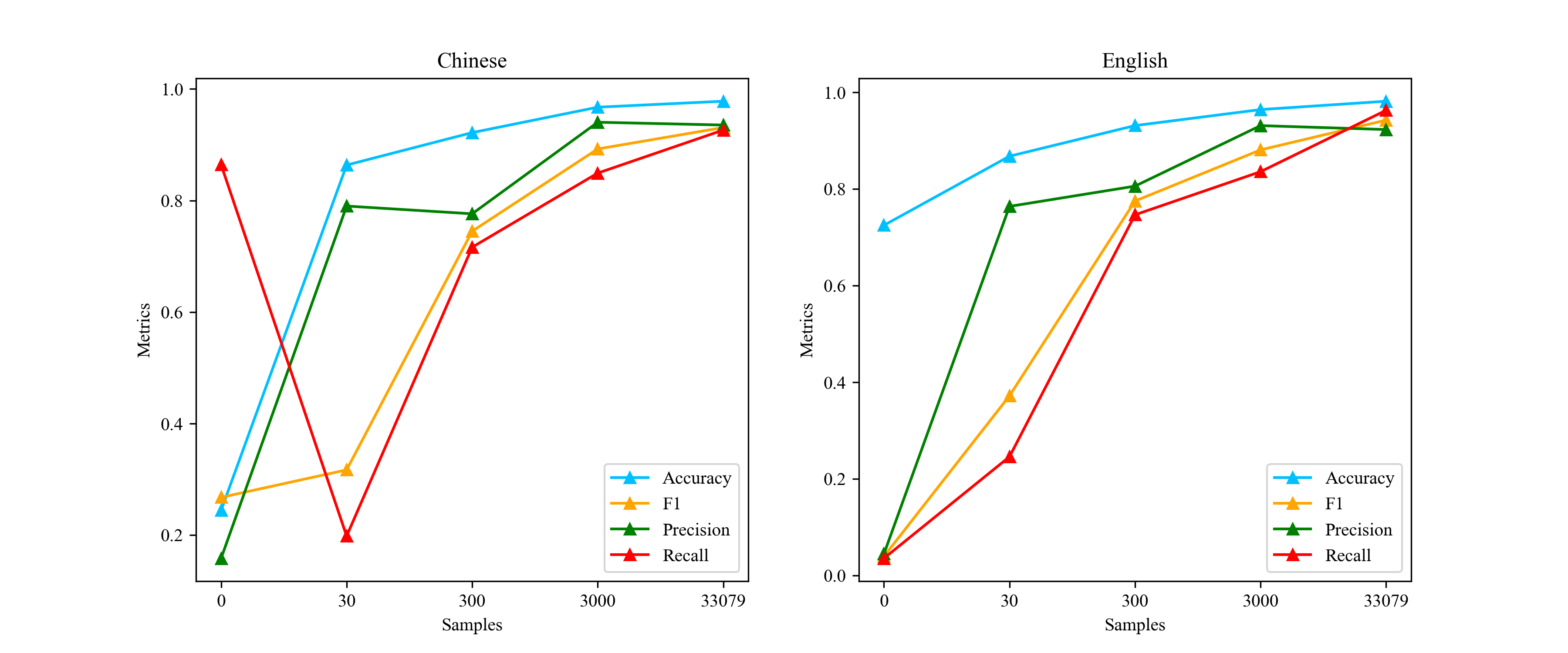}
    \caption{Changes of 4 metrics under different sample numbers.}
    \label{fig:samples}
\end{figure*}

One of the key findings in our main experiment is that the zero-shot prompt evaluation method performs poorly on both datasets.
Regardless of the backbone PLM used, it falls far behind the baseline methods and the latter two types of methods.
However, it has been mentioned in previous work \cite{brown2020language} that PLMs possess a certain level of few-shot learning capability.
Hence, we further conducted few-shot testing on the DuRecDial 2.0 dataset using the prompt learning method with BERT-base as the backbone.
To ensure sufficient learning under few-shot conditions, we adjusted the experimental settings for 30-shot and 300-shot conditions to train for 100 and 10 times the number of epochs, respectively.
The 3000-shot condition remained consistent with the main experiment.

\begin{table*}[htb]
\centering
\caption{Experimental results of hard prompt learning with BERT-base using the different samples on DuRecDial 2.0 dataset.}
\label{tab:samples}
\resizebox{\linewidth}{!}{
\begin{tabular}{c|cccc|cccc}
  \toprule
  \multirow{2}{*}{Samples} & \multicolumn{4}{c|}{English} & \multicolumn{4}{c}{Chinese} \\
                   & Accuracy  & Precision & Recall & F1
                   & Accuracy  & Precision & Recall & F1 \\
  \midrule
  0
                   & 0.7244 	& 0.0444 	& 0.0353 & 0.0393
                   & 0.2444 	& 0.1583 	& 0.8641 & 0.2676 \\
  30
                   & 0.8673 	& 0.7635 	& 0.2455 & 0.3715
                   & 0.8635 	& 0.7900 	& 0.1982 & 0.3169 \\ 
  300
                   & 0.9307 	& 0.8054 	& 0.7461 & 0.7746 
                   & 0.9217 	& 0.7761 	& 0.7162 & 0.7449 \\
  3000
                   & 0.9638 	& 0.9306 	& 0.8353 & 0.8804
                   & 0.9673 	& 0.9403 	& 0.8491 & 0.8924 \\
  33079 (Full samples)
                   & 0.9811 	& 0.9225 	& 0.9623 & 0.9420 
                   & 0.9780 	& 0.9353 	& 0.9263 & 0.9308 \\
  \midrule
  30 (balanced)
                   & 0.7677 	& 0.3902 	& 0.8078 & 0.5262 
                   & 0.7235 	& 0.3323 	& 0.7246 & 0.4557 \\ 
  300 (balanced)
                   & 0.9276 	& 0.7422 	& 0.8377 & 0.7871
                   & 0.8996 	& 0.6532 	& 0.7916 & 0.7158 \\
  \bottomrule
\end{tabular}
}
\end{table*}

As illustrated in Fig.~\ref{fig:samples} and the first part of Table~\ref{tab:samples}, the increase in data samples results in improved performance in recommendability identification.
Notably, under the 30-shot condition, BERT-base's accuracy has reached a relatively stable level.
Under the 300-shot condition, all four metrics have achieved higher levels than the zero-shot scenario.

Considering that only four positive instances exist in the 30-shot data, it is challenging for the model to learn broad recommendability features from these four positive instances.
Therefore, we conducted further experiments to balance the positive and negative instances under the 30-shot and 300-shot conditions (with an equal number of positive and negative samples).
As shown in the second part of Table~\ref{tab:samples}, BERT-base achieved a significantly better recall rate and F1 scores under the 30-shot condition. 
These experiments demonstrate PLMs' few-shot learning capability for recommendability identification under appropriate positive sample ratios.

\begin{table*}[htb]
\caption{Experimental results of hard prompt learning with BERT-base using different manual templates on DuRecDial 2.0 dataset.}
\label{tab:template}
\resizebox{\linewidth}{!}{
\begin{tabular}{cc|cccc|cccc}
  \toprule
  \multirow{2}{*}{Method Type} & \multirow{2}{*}{Template ID} & \multicolumn{4}{c|}{English} & \multicolumn{4}{c}{Chinese} \\
  &
                   & Accuracy  & Precision & Recall & F1
                   & Accuracy  & Precision & Recall & F1 \\
  \midrule
  \multirowcell{2}{Zero-shot \\ Prompt Evaluation} & 1        
                   & 0.7244 	& \textbf{0.0444} 	& \textbf{0.0353} & \textbf{0.0393}
                   & 0.2444    & \textbf{0.1583}    & \textbf{0.8641} & \textbf{0.2676} \\
        & 2        & \textbf{0.7658}    & 0.0233    & 0.0114 & 0.0153 
                   & \textbf{0.4090} 	& 0.1516 	& 0.5874 & 0.2410 \\
  \midrule
  \multirow{2}{*}{Hard Prompt Learning} & 1        
                   & \textbf{0.9811} 	& \textbf{0.9225} 	& 0.9623 & \textbf{0.9420}
                   & \textbf{0.9780} 	& 0.9353 	& \textbf{0.9263} & \textbf{0.9308} \\
        & 2        & 0.9809    & 0.9148    & \textbf{0.9707} & 0.9419
                   & 0.9745 	& \textbf{0.9705} 	& 0.8665 & 0.9155 \\
  \bottomrule
\end{tabular}
}
\end{table*}

\subsubsection{Influence of Manual Template}

Considering the impact of manual templates on the model's results, we designed a second template (including both Chinese and English versions) for the DuRecDial 2.0 dataset.
We conducted comparative experiments on the zero-shot prompt evaluation and hard prompt learning methods, utilizing BERT-base as the backbone.

As illustrated in Table~\ref{tab:template}, there are specific differences in performance between Template 1 (the manual template used in the main experiment) and Template 2 under the zero-shot evaluation setting.
However, under the hard prompt learning setting, the metrics of the two manual templates are very close.

Overall, Template 1, used in the main experiment, performed better.
However, it is essential to acknowledge that the construction of manual templates can influence experimental results.
Furthermore, there is still plenty of room for optimizing the performance of manual template construction in the recommendability identification task.

\subsubsection{Influence of Soft Prompt Length}

To show the effects of different soft prompt lengths on the P-tuning method, we have conducted a comparative experiment on the DuRecDial 2.0 English dataset with BERT-base as the backbone.

\begin{figure}
    \centering
    \includegraphics[width=0.75\linewidth]{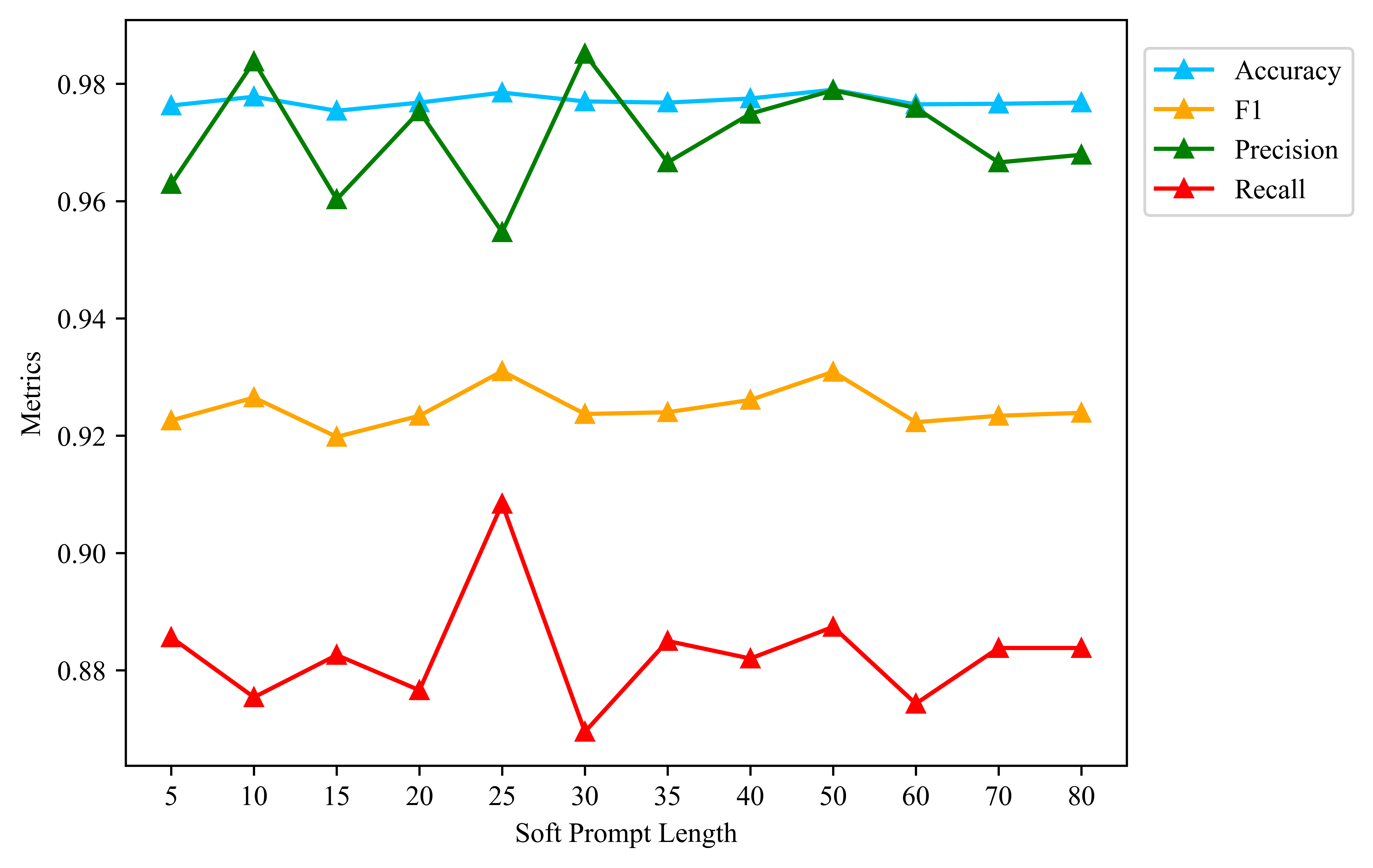}
    \caption{Performance comparison across different soft prompt lengths.}
    \label{fig:prefix}
\end{figure}

The results of all metrics as soft prompt length varies are presented in Fig.~\ref{fig:prefix}.
The results of this figure show that the model's accuracy and F1 scores do not change significantly as the length of the soft prompt changes.
This suggests the soft prompt tuning method performs consistently well in the recommendability identification task.
It also confirms the beneficial role of fewer training parameters in ensuring experimental stability.

\subsection{Discussion and Limitation}


This work primarily introduces the recommendability identification task and reveals the potential of leveraging PLMs through experiments.
While we have achieved some encouraging results, there are still limitations:

First, constrained by computation resources, our experiments could only cover a limited number of LLMs and conduct prompt evaluation under the zero-shot condition.
In the future, further experiments can be conducted with a broader range of LLMs.

Second, due to the limitations on the encoding length of various PLMs, we did not leverage rich user profiles and context information to design our manual templates.
This area is also worth exploring in future research.

Third, JDDCRec, consisting of 17k utterances and 1k conversations obtained through manual annotation, represents our first step in studying recommendability.
We hope that JDDCRec can be further explored and expanded, encouraging more researchers to construct larger-scale datasets.
Additionally, automated data collection from real-life scenarios is an exciting and meaningful research direction as well.
\section{Conclusion}



Our work took the first step in recommendability identification in conversations, integrating the fields of conversation and recommender systems.
We have achieved encouraging results and showed the potential of incorporating recommendation capabilities into conversation systems.
The findings emphasize the importance of minimizing user disruption, particularly regarding recommendation behavior, to enhance the overall user experience.

Further improvements can be made in refining the recommendability identification models and exploring novel techniques.
Additionally, it will be crucial to develop real-time dynamic identification mechanisms that adapt to users' changing needs and preferences.
By trade-off between providing relevant recommendations and respecting user needs, we can apply conversational search and recommendation systems in more scenarios.

In summary, our work highlights the significance of giving recommendations at the right time in conversations and paves the way for future research in this domain.
By considering user needs and minimizing disruptions, we can improve user satisfaction and the overall performance of conversation systems.

\subsubsection{\ackname} This work is supported by the Natural Science Foundation of China (Grant No. U21B2026, 62372260) and Quan Cheng Laboratory (Grant No. QCLZD202301).
%
%
%
\bibliographystyle{splncs04}
\bibliography{reference}
%




\end{document}